\documentclass[10pt,a4paper]{article}

\usepackage{amssymb}			
\usepackage{amsmath}			
\usepackage{graphicx}			
\usepackage[latin1]{inputenc}	
\usepackage{hyperref}			
\usepackage[usenames,dvipsnames,svgnames,table]{xcolor}						
\hypersetup{colorlinks=true, citecolor=Blue, linkcolor=Blue, urlcolor=Blue}				

\usepackage{graphicx}
\usepackage{color}

\usepackage{amsmath,amsfonts,amssymb,amsthm,epsfig,epstopdf,url,array}
\usepackage{float}

\usepackage{wrapfig}
\usepackage{cancel}

\newtheorem{thm}{Theorem}[section]

\newcommand{\R}{{\mathbb R}}
\newcommand{\q}{{\quad}}

\newcommand{\be}{\begin{equation}}
	\newcommand{\ee}{\end{equation}}

\def\be{\begin{equation}}
	\def\ee{\end{equation}}

\newcommand{\RomanNumeralCaps}[1]

\begin{document}
	
	\title{Squire's theorem for nonlinear monotone energy stability}
	\author{Giuseppe Mulone\footnote{Universit\`{a} degli Studi di Catania (retired), Dipartimento di Matematica e Informatica, Viale Andrea Doria 6, 95125 Catania, Italy,  giuseppe.mulone@unict.it}}
	\date{}
	\maketitle

\begin{abstract}
The Squire's theorem holds for parallel shear flows governed by the linearized Navier-Stokes equations. Squire writes ``For the study of the stability of flow between parallel walls it is sufficient to confine attention to disturbances of two-dimensional type", Squire \cite[p. 627]{Squire1933}. Instead, for  nonlinear Navier-Stokes system it is supposed that the theorem does not hold in general (see Drazin and Reid \cite[pp. 429--430]{DrazinReid2004}).

Here we prove that the Squire theorem holds also for nonlinear monotone energy stability of parallel shear flows that include Couette and Poiseuille flows between parallel walls.
	
%
%
\end{abstract}

%

\parindent 0pt



\section{Introduction}
\label{sec:intro}

Squire, \cite{Squire1933}, considers the steady two-dimensional flow of an incompressible viscous fluid upon which a \textit{small} disturbance is superimposed (linearized Navier-Stokes equations). In the summary of his paper, he  writes: ``{The stability of the flow of a viscous fluid between parallel walls for three-dimensional disturbances is discussed. A fourth-order differential equation is
derived and it is shown that, if any velocity profile is unstable for a particular value of Reynolds' number, it will be unstable at a lower value of Reynolds' number for two-dimensional disturbances}". This is now called \textit{Squire theorem}, see Drazin and Reid \cite{DrazinReid2004}: \textit{``to each unstable three-dimensional disturbance there corresponds a more unstable
two-dimensional one".}

Squire's theorem has also been proved for many other fluids including flows in magnetohydrodynamics and in viscoelastic fluids, see \cite{Stuart.1954} - \cite{daSilva.2022}.

Drazin and Reid in their monograph,  \cite[pp. 429--430]{DrazinReid2004}, write in the case of nonlinear theory:

``{Orr (1907) further assumed that two-dimensional perturbations, i.e. the least eigenvalue of equations (53.21) can be obtained as the least eigenvalue of equation (53.22). Although Squire's theorem provides a justification for this assumption in the case of linear theory, it is not applicable to equations (53.21) and the assumption is false in general. 
\\
The determination of the least eigenvalue of equations (53.21) is clearly a formidable problem in general and results are known for only a few flows. In the case of Couette flow with $U(z)=z$  $(-1\le z\le 1)$, however, Joseph (1966) has proved that the least eigenvalue of equations (53.21) is still associated with a two-dimensional disturbance but one which varies only in the $yz$-plane, i.e. the perturbed flow consists of rolls whose axes are in the direction of the basic flow.}"

In the nonlinear case, studying the nonlinear stability with the Lyapunov method and choosing the energy of the perturbation as the Lyapunov function, the critical energy Reynolds number is obtained by determining the  maximum of a functional ratio obtained from the Reynolds-Orr equation, Orr \cite{Orr1907}. Orr assumed (without proving it) that the maximum is obtained on the perturbations to the velocity field which satisfy no-slip boundary conditions,  are periodic, divergence-free, and they are two-dimensional spanwise. Joseph, \cite{Joseph.1966}, \cite{Joseph.1976} instead assumed that the maximum is attained on streamwise perturbations (2.5-dimensional perturbations).

Busse, \cite{Busse1972} writes ``SQUIRE'S Theorem does not hold for the Euler equations determining the energy stability limit ${\rm Re}_E$". 

Despite these results, this problem was taken up by Falsaperla et al. \cite{FalsaperlaGiacobbeMulone2019} - \cite{FalsaperlaMulonePerrone2022}. 
In particular, in \cite{FalsaperlaMulonePerrone2022}, the authors  conjectured that the search  for the maximum should be performed on a subspace of the space of kinematically admissible perturbations, the spanwise perturbations, as Orr assumed.

Recently, Mulone \cite{Mulone.2023} proved that the conjecture of \cite{FalsaperlaMulonePerrone2022} is true.
A consequence of this proof is that Squire's theorem holds also for the nonlinear monotone energy stability: \textit{to obtain the minimum critical nonlinear-energy Reynolds number it is sufficient to consider only two-dimensional spanwise disturbances}.

The plan of the paper is as follows. 
In Sec. 2 we write the non-dimensional perturbation equations of basic laminar flows between two horizontal planes with no-slip boundary conditions.
In Section 3, we recall the Squire theorem for linear Navier-Stokes system.
In Sec. 4 we study monotone nonlinear energy stability and prove that the  Squire theorem holds for nonlinear monotone energy stability. In Sec. 5  we  make  a conclusion.

\section{Laminar flows between two parallel planes}
Given  a reference frame $Oxyz$, with unit vectors ${\bf i},{\bf j}, {\bf k}$, consider the (non-dimensional) layer $\mathcal D = \R^2 \times [-1, 1]$  of thickness $2$ with horizontal coordinates $x,y$ and vertical coordinate (orthogonal to the layer) $z$.

The basic flows which are  solutions of the  stationary Navier-Stokes equations, \cite{DrazinReid2004}, are 
characterized by the functional form
\begin{align}
	\label{basic}
	{\bf U}= (	f(z),0,0) =  f(z) {\bf i},
\end{align}
where ${\bf U}$ is the velocity field. 
The function $f(z) : [-1, 1] \to \mathbb R$ is assumed to be sufficiently smooth and it is called the shear profile (base or mean flow). The velocity filed is written in a non-dimensional form. To non-dimensionalize the Navier-Stokes equations and the gap of the layer we use a Reynolds number ${\rm Re}$ based on the average shear and half gap $d$ (see \cite{FalsaperlaGiacobbeMulone2019}).

In particular, for fixed velocity at the boundaries $z=\pm 1$, we consider two well known profiles: \textit{Couette} flow $f(z)=z$, and \textit{Poiseuille} flow $f (z) = 1-z^2$.	

\subsection{Perturbation equations}
The perturbation equations to the plane parallel shear flows, in non-dimensional form, are 
\begin{align}
	\label{Couette-gen}
	\left\{ \begin{array}{l}
		u_t = -  {\bf u}\!\cdot\!\nabla u+ {\rm Re}^{-1} \Delta u -  (f u_x+f' w)- p_x\\[5pt]
		v_t = -  {\bf u}\!\cdot\!\nabla v+ {\rm Re}^{-1} \Delta v -  f v_x - p_y\\[5pt]
		w_t = -  {\bf u}\!\cdot\!\nabla w+  {\rm Re}^{-1}\Delta w -  f w_x- p_z\\	[5pt]
		u_x+v_y+w_z=0 .\\	
	\end{array}  \right.
\end{align}
In \eqref{Couette-gen} ${\bf u}$ is  the perturbation velocity field and $p$ the perturbation to the pressure field. The velocity field {\bf u}  has components  $(u, v, w)$ in the directions $x,y,z$, respectively.
Here we use the symbols $g_x$ as $\frac {\partial g} {\partial x}$, $g_t$ as $\frac {\partial g} {\partial t}$, etc., for any function $g$.
To system (\ref{Couette-gen}) we append the \textit{rigid} (no-slip) boundary conditions $${\bf u}(x,y,\pm 1,t)=0, \q (x,y,t) \in  \R^2 \times (0, +\infty), $$ and the initial condition
$${\bf u}(x,y,z,0)= {\bf u_0}(x,y,z), \q {\rm in}  \q\mathcal D,$$
with ${\bf u_0}(x,y,z)$ solenoidal vector which vanishes at the boundaries.

We recall that the streamwise (or longitudinal) perturbations $(u.v.w), p$ are those fields which do not depend on $x$; the spanwise (or transverse) perturbations are those fields which do not depend on $y$. In particular, we have two-dimensional spanwise perturbations if also $v=0$.

\subsection{Linear stability/instability, the Squire's theorem}
Assume that both ${\bf u}$ and $\nabla p $ are $x,y$-periodic with periods $2\pi/a$ and $2\pi/b$ in the $x$ and $y$ directions, respectively,  with wave numbers $( a, b) \in \R^2_+$  . In the following, it suffices therefore to consider functions over the periodicity cell 
$$\Omega= [0, \frac{2\pi}{a}]\times [0, \frac{2\pi}{ b}] \times [- 1, 1] .$$

Linear stability/instability is obtained by studying the linearized system neglecting the nonlinear terms in \eqref{Couette-gen}.

We recall that, for laminar flows between parallel planes,  Romanov \cite{Romanov1973}  proves that \textit{Couette} flow is \textit{linearly stable} for \textit{any Reynolds} number.   Orszag \cite{Orszag1971} proves that \textit{Poiseuille} flow is \textit{unstable} for any Reynolds number bigger that $5772$. In the last case, Orszag assumes two-dimensional spanwise disturbances (in agreement with the results of Orr and Squire) and solves the Orr-Sommerfeld equation, see Drazin and Reid \cite[p. 156]{DrazinReid2004}.

The Squire's theorem, \cite{Squire1933}, holds for the linearized system: the most destabilizing perturbations are two-dimensional spanwise  perturbations.

\section{Nonlinear monotone energy stability}
As the basic function space, we take $L_2(\Omega)$, which is the space of square-summable functions in $\Omega$ with the scalar product denoted by
$$(g,h) = \int_0^{\frac{2\pi}{a}} \int_0^{\frac{2\pi}{ b}} \int_{-1}^1 g(x,y,z) h(x,y,z) dxdydz, $$ 
and the corresponding  norm $\Vert g \Vert = (g,g)^{1/2}.$

The \textit{ nonlinear monotone  energy stability} is based on the \textit{Reynolds-Orr energy equation}, \cite{Reynolds1895}
\begin{align}
	\label{Energy}
	\dot E= -(f'w,u) - {\rm Re}^{-1} [\Vert \nabla u \Vert^2+\Vert \nabla v \Vert^2+\Vert \nabla w \Vert^2], 
\end{align}
where 
\be \label{EN-EQ}
E(t) = \dfrac{1}{2}[\Vert u \Vert^2 +  \Vert v \Vert^2 + \Vert w \Vert^2 ] 
\ee
is the (density) \textit{ energy} of disturbance.

We first note that Lorentz \cite{Lorentz.1907}  made this observation (see Lamb \cite{Lamb.1924} , p. 640) : ``One or two consequences of the energy equation may be noted. In the first place, the relative magnitude of the two terms on the right-hand side  is unaffected if we reverse the signs of $u, v, w$, or if we multiply them by any constant factor. The stability of a given state of mean motion should not therefore depend on the \textit{scale} of the disturbance".
This \textit{scale invariance} property in particular means that the $-(f'u,w)$ term has a positive or negative sign and cannot change by exchanging the sign of only one of the two components $u$ and $w$. 

For perturbations with  $ -(f'w,u) \le 0$ and $\Vert \nabla {\bf u} \Vert>0$, then $\dot E < - C_0 E$, with $C_0$ a positive (Poincaré) constant. If instead, $-(f'u,w)>0$, from \eqref{Energy}, we easily have


	\begin{eqnarray}
	\begin{array}{l}\label{ineq-}
		\dot E=  \left(\dfrac{-(f'w,u)}{\Vert \nabla u \Vert^2+\Vert \nabla v \Vert^2+\Vert \nabla w \Vert^2} - \dfrac{1}{{\rm Re}}\right)\Vert \nabla {\bf u} \Vert^2  
		\le\left(m - \dfrac{1}{{\rm Re}}\right)\Vert \nabla {\bf u} \Vert^2 ,
	\end{array}
\end{eqnarray}
where
\begin{eqnarray} \label{max-iniz}
	\dfrac{1}{{\rm Re}_E} = m= \max_{\cal S} \dfrac{-(f'w,u)}{\Vert \nabla u \Vert^2+\Vert \nabla v\Vert^2+\Vert \nabla w \Vert^2}, 
\end{eqnarray}
$\cal S$ is the space of the   {\textit{kinematically admissible fields}}  
\begin{eqnarray} \nonumber
	\begin{array}{l}
		\label{spaceS}
		{\cal S}= \{u, v, w \in H^1(\Omega), \; u=v=w=0 \q \hbox{on the boundaries,}\\[3mm] \hbox{ periodic in \textit{x}, and \textit{y},}  \q u_x+v_y+w_z=0,\q  \Vert \nabla {\bf u} \Vert>0\},
	\end{array}
\end{eqnarray}
with $H^1(\Omega)=W^{1,2}(\Omega)$ the Sobolev space: the subset of functions ${\displaystyle h} \in {\displaystyle L^{2}(\Omega )}$ such that ${\displaystyle h}$ and its weak derivatives up to order ${\displaystyle 1}$ have a finite $L^2$-norm.

Assuming $-(f'w,u)>0$, consider the functional ratio	
 \[\label{p1} 
	{\cal F}(u,v,w)= \frac{-(f'w,u)}{\Vert \nabla u \Vert^2+\Vert \nabla v\Vert^2+\Vert \nabla w\Vert^2}  \]
		in ${\cal S}$. 
For any 	$(u,v,w)$ in $\cal S$ satisfying the scale invariance property, we have
\begin{align}
	\nonumber 
		{\cal F}(u,v,w) \le \frac{-(f'w,u)}{\Vert \nabla u\Vert^2+\Vert \nabla w\Vert^2} 
		\le \max_{\cal S} \frac{-(f'w,u)}{\Vert \nabla u\Vert^2+\Vert \nabla w\Vert^2}  .
\end{align}
In \cite{Mulone.2023} it has been proved that
\begin{align}\label{prefin}
\max_{\cal S} \frac{-(f'w,u)}{\Vert \nabla u\Vert^2+\Vert \nabla w\Vert^2} =
\max_{\cal S} 	{\cal F}(u,0,w),
\end{align}
with ${\bf u}= (u(x,y,z), 0, w(x,,y,z))$ which do not depend on $y$ and  satisfy the scale invariance property (note that the scale invariance property is not verified for streamwise perturbations used by Joseph \cite{Joseph.1966} - \cite{Joseph.1976}). To obtain \eqref{prefin} it is sufficient to solve the Euler-Lagrange equations of the maximum problem and take into account  the boundary conditions, the solenoidality of the velocity field, the periodicity, and the invariance of scale of the energy equation (cf. \cite{Mulone.2023}).
The Euler-Lagrange equations of the maximum \eqref{prefin} are: $-f' w +2 \bar m \Delta u = \lambda_x, 0= \lambda_y, -f' u +2 \bar m \Delta w =\lambda_z$, where $\bar m$ is the maximum, and $\lambda$ is a Lagrange multiplier. Searching solutions $(u,v,w,\lambda)= \bar H(z) e^{i(ax+by)}$, with $a$ and $b$ wave numbers and $\bar H(z)= \bar u(z), \bar v(z), \bar w(z), \lambda(z)$ respectively, we have: $b=0$. Therefore, $u,v,w, \lambda$ are functions of $x$ and $z$, and $v_y=0$, $u_x+w_z=0$. Since the space of fields $(u(x,z),v(x,z),w(x,z))$ in $\cal S$, with $u_x+w_z=0$,  is a subspace of $\cal S$, the two maxima: the maximum \eqref{max-iniz} and the maximum on the first side of equation \eqref{prefin} are equal. Hence, $v_x=v_z=0$. These equations and $v_y=0$ imply $v=0$ for the maximizing field. Note that if $\lambda =0$, then from the Euler-Lagrange equations and the boundary conditions we would have $u=w=0$. In this case it should be noted that the perturbation $(0, v(x,z),0)$ is always stabilizing.

Therefore, we have
\begin{align}\label{finale}
\max_{\cal S} {\cal F}(u,v,w) = \max_{\cal S} 	{\cal F}(u(x,z),0,w(x.z)),
\end{align}
because the space of fields $(u,0,w)$ in $\cal S$, with $u_x+w_z=0$,  is a subspace of $\cal S$.

The Euler-Lagrange equations of the two maxima problems \eqref{max-iniz} and \eqref{prefin}, after eliminating the Lagrange multipliers, become

- for the $\max_{\cal S} {\cal F}(u,v,w)$  :
\begin{align}
	\label{el-S}
	\begin{cases}
	f'(\zeta_y+2w_{xz})+f''w_x+2m \Delta \Delta w   =0\\
		f'w_y+2m\Delta \zeta=0 ,
	\end{cases}
\end{align}
where $\zeta= v_x-u_y$ is the third component of vorticity. The boundary conditions are $w=w_z=\zeta=0$ on the boundaries.

- for the $\max_{{\cal S}}  {\cal F}(u,0,w)$:
\begin{align} 
	\label{El-O} 2f' w_{xz}+f''w_x+2m \Delta \Delta w   =0, 
\end{align}
with boundary conditions $w=w_z=0$.

It is easy to verify that the last equation and the boundary conditions may be derived immediately from system \eqref{el-S} and the boundary conditions by putting $v=0$ and $\partial /\partial y \equiv 0$. This is exactly what happens for the equation (26) of Orr \cite{Orr1907}.

Therefore, we have

\begin{thm}
	The critical nonlinear energy Reynolds number ${\rm Re}_E$ is given by solving the Orr's equation 
\begin{align} 
	\label{Orr-eq}
	{\rm Re}_E (f'' w_x+2f' w_{xz})+ 2\Delta \Delta w=0.	
\end{align}	with b.c. $w=w_z=0$, and  $\Delta = \dfrac{\partial^2}{\partial x^2} + \dfrac{\partial^2}{\partial z^2}$.
\end{thm}
\begin{thm}\label{t31}
	\textit{Assuming ${\rm Re} <{{\rm Re}_E}$, the  basic shear flow (\ref{basic}) is nonlinear monotone exponentially stable according to the  energy:
		\be\label{mon-stab} E(t) \le E(0) \exp\left\{{\dfrac{\pi^2}{2}\left( \dfrac{1}{{\rm Re}_E} - \dfrac{1}{{\rm Re}}\right)t}\right\}, \q \forall t \ge 0.
		\ee 
} \end{thm}

%
\begin{thm}
	\textbf{(Squire Theorem} for nonlinear energy stability). 
	
	\textit{The least stabilizing perturbations in the energy norm \eqref{EN-EQ} are two-dimensional perturbations $(u,0,w)$ which do not depend on $y$, the spanwise  perturbations.
	}
\end{thm}

\section{Conclusion}

We have proved that, in the case of the classical flows of Couette and Poiseuille between parallel walls, Squire's theorem holds also in the nonlinear energy stability:
for the study of the nonlinear energy stability of flow between parallel walls it is sufficient to confine attention to disturbances of two-dimensional spanwise perturbations. This result was obtained by comparing two maximum functional problems.

We note that Busse \cite{Busse1972} uses the correct Reynolds-Orr energy equation, but defines an incorrect functional ratio (see \cite{Busse1972}, formula (2)), precisely \textit{replacing $-(f' u,w)$ with its absolute value}. This leads to the erroneous result of his manuscript and of Joseph's paper: they only get sufficient nonlinear stability conditions and underestimated the critical Reynolds number.  For instance, in the case of Couette, the nonlinear Reynolds number they obtain is $20.6$ instead of the exact critical Reynolds value $44.3$ obtained by Orr.

Finally, we note that the Squire's theorem may be also proved to other fluid flows including nonlinear flows in magnetohydrodynamics and to viscoelastic fluids, and possibly to flows between parallel planes with other boundary conditions.

\vskip .4cm

\textbf{Statements and Declarations}

I declare that I have no competing financial and / or non-financial interests.
\vskip.4cm


\vskip .4cm
{\bf Acknowledgements}\\
{The research that led to the present paper was partially supported by the following Grants: 2017YBKNCE of national project PRIN of Italian Ministry for University and Research, PTR DMI-53722122146 "ASDeA" of the University of Catania. We also thank the group GNFM of INdAM for financial support.}

\bibliographystyle{jfm}

\end{document}